# Application of Quantum Convolutional Neural Networks for MRI-Based Brain Tumor Detection and Classification

Sugih Pratama Nugraha[1], Ariiq Islam Alfajri[1], Tony Sumaryada[1], Duong Thanh Tai[2,*], Nissren Tamam[3], Abdelmoneim Sulieman[4,5], Sitti Yani[1,*]

[1]Department of Physics, Faculty of Mathematics and Natural Sciences, IPB University. Bogor, Indonesia. Email: sittiyani@apps.ipb.ac.id; ariiq.i.alfajri@gmail.com; tsumaryada@apps.ipb.ac.id; Sugihpratamas@gmail.com

[2]Department of Medical Physics, Faculty of Medicine, Nguyen Tat Thanh University, Viet Nam. Email: dttai@ntt.edu.vn

[3]Department of physics, College of Science, Princess Nourah bint Abdulrahman University, P.O Box 84428,Riyadh 11671,Saudi Arabia. Email: nissrentamam@gmail.com

[4]Department of Radiological Sciences, College of Applied Medical Sciences, King Saud Bin Abdulaziz University for Health Sciences, Al Ahsa, Saudi Arabia. Email: abdelmoneim_a@yahoo.com

[5]Imaging Physics Section, Research and Innovation Department, King Faisal Specialist Hospital & Research Center, Riyadh, Saudi Arabia

Corresponding author: Dr. Sitti Yani (sittiyani@apps.ipb.ac.id); Dr. Duong Thanh Tai (dttai@ntt.edu.vn); Prof. Abdelmoneim Sulieman (abdelmoneim_a@yahoo.com);

**Abstract:**

This study explores the application of Quantum Convolutional Neural Networks (QCNNs) for brain tumor classification using MRI images, leveraging quantum computing for enhanced computational efficiency. A dataset of 3,264 MRI images, including glioma, meningioma, pituitary tumors, and non-tumor cases, was utilized. The data was split into 80% training and 20% testing, with an oversampling technique applied to address class imbalance. The QCNN model consists of quantum convolution layers, flatten layers, and dense layers, with a filter size of 2, depth of 4, and 4 qubits, trained over 10 epochs. Two models were developed: a binary classification model distinguishing tumor presence and a multiclass classification model categorizing tumor types. The binary model achieved 88% accuracy, improving to 89% after data balancing, while the multiclass model achieved 52% accuracy, increasing to 62% after oversampling. Despite strong binary classification performance, the multiclass model faced challenges due to dataset complexity and quantum circuit limitations. These findings suggest that QCNNs hold promise for medical imaging applications, particularly in binary classification. However, further refinements, including optimized quantum circuit architectures and hybrid classical-quantum approaches, are necessary to enhance multiclass classification accuracy and improve QCNN applicability in clinical settings.

## 1. Introduction

Brain tumors pose a formidable global health challenge, constituting the majority of primary central nervous system (CNS) neoplasms and significantly contributing to cancer-related morbidity and mortality. Incidence rates exhibit regional variation, yet projections estimate tens of thousands of new cases and fatalities annually in the United States alone, with hundreds of thousands diagnosed worldwide each year. Clinically, brain tumors present with symptoms such as headaches and neuropsychiatric disturbances, including personality changes, memory impairment, and anxiety disorders [1–3]. They are broadly classified into primary tumors, originating within the brain, and secondary (metastatic) tumors, arising from malignancies in other organs. Among primary tumors, gliomas, meningiomas, and pituitary adenomas predominate [2,4]. The dynamic epidemiology of brain tumors—marked by shifting incidence trends and distinct impacts on populations such as adolescents and young adults—underscores the urgent need for innovative diagnostic technologies.

Magnetic Resonance Imaging (MRI) is the cornerstone of neuro-oncological imaging, offering exceptional soft-tissue contrast vital for delineating brain anatomy and characterizing tumor attributes, including size, location, and morphology [5,6]. Beyond structural insights, MRI informs diagnosis, prognosis, and treatment planning [7]. However, traditional MRI interpretation relies heavily on qualitative visual assessment by radiologists, a process prone to inter-observer variability and time-intensive, especially with large datasets [7,8]. This MRI images can be evaluated and classified using machine learning and transfer learning [9–11]. Moreover, conventional MRI techniques face inherent limitations, such as challenges in precisely defining the boundaries of infiltrative tumors that extend beyond contrast-enhanced regions—often due to variable blood-brain barrier disruption [12]and distinguishing true tumor progression from treatment-related changes like pseudoprogression. These shortcomings highlight the demand for automated, objective computational approaches capable of discerning subtle, clinically significant patterns in complex multi-parametric MRI data that may elude visual inspection.

In recent years, classical deep learning (DL) techniques, notably Convolutional Neural Networks (CNNs), have emerged as robust tools for automating medical image analysis, including brain MRI scans [13]. CNNs, a subset of machine learning, emulate neural processes to detect intricate patterns through interconnected nodes designed to mimic biological neurons [14,15]. These networks have demonstrated remarkable efficacy, achieving classification accuracies exceeding 90% in both binary and multiclass brain tumor detection tasks[16–18]. Despite their success, classical CNNs are computationally intensive, prompting exploration into advanced paradigms like Quantum Machine Learning (QML). QML leverages quantum computing principles to tackle complex computational challenges, offering potential advantages in speed and precision. Quantum Convolutional Neural Networks (QCNNs), a quantum-inspired evolution of CNNs, integrate convolutional, subsampling, and fully connected layers, with the convolutional layer employing quantum circuits and application-specific quantum gates [19–23]. Several researchers implemented QCNN models on Quantum's TensorFlow platform using the MNIST and Fashion MNIST datasets successfully classifying these datasets with accuracy above 90% ( [19,22–24]. In addition, it was also found that QNN can shorten the computation time with better accuracy in classification [25]. Quantum Machine Learning (QML) is also used in the field of high energy physics. Several QCNN setups are compared by varying the convolutional circuit, type of encoding, loss function, and batch sizes. The results obtained show that QCNNs with appropriate setups can produce better accuracy compared to classical CNNs especially under the condition of convolutional blocks with a smaller number of parameters [21]. Another study also investigated the effect of quantum filter structure, filter order and filter parameters on QCNN performance in multiclass classification. The results obtained show that an appropriate quantum structure can significantly improve model performance [26]. Classification of breast cancer images using Quantum Convolutional Neural Network (QCNN) was also carried out resulting in quite good accuracy, above 75% [27,28]. Despite these advancements, the application of QCNNs to brain MRI classification remains limited. The application of QCNNs, particularly in medical imaging, remains in its exploratory stages [29].

Current research often relies on classical simulations or hybrid quantum-classical architectures due to the constraints of existing Noisy Intermediate-Scale Quantum (NISQ) hardware. A definitive demonstration of practical "quantum advantage" over state-of-the-art classical methods for real-world medical data analysis is yet to be established and is an active area of investigation, complicated by questions surrounding the classical simulability of some QCNN models and benchmark tasks. While previous QCNN studies have primarily focused on generic datasets such as MNIST or non-medical benchmarks, few have rigorously evaluated QCNN performance on clinical brain MRI images using a direct comparison with conventional CNNs. To bridge this gap, this work proposes a compact four-qubit QCNN specifically designed for brain tumor detection and classification. The model incorporates an entangled Pauli-Z residual filter and a hybrid variational training strategy that jointly optimizes quantum and classical layers, enabling a significant reduction in trainable parameters while maintaining competitive accuracy. By analyzing feature embeddings and testing different circuit configurations, this study provides insights into the practical trade-offs between accuracy, efficiency, and noise resilience for real-world neuroimaging tasks on near-term quantum devices. Therefore, in this research, a Quantum Convolutional Neural Network (QCNN) is developed to classify MRI images. Classification is divided into two types, namely binary class (tumor and non-tumor) and multiclass (meningioma, glioma, pituitary, and non-tumor).

## 2. Materials and Methods
## 2.1. Data Preprocessing

The dataset comprises brain MRI scans from subjects diagnosed with glioma, meningioma, and pituitary tumors, alongside scans from individuals with non-tumor brain conditions. This dataset was sourced from Kaggle [30]. Representative MRI images across these categories are presented in Figure 1.

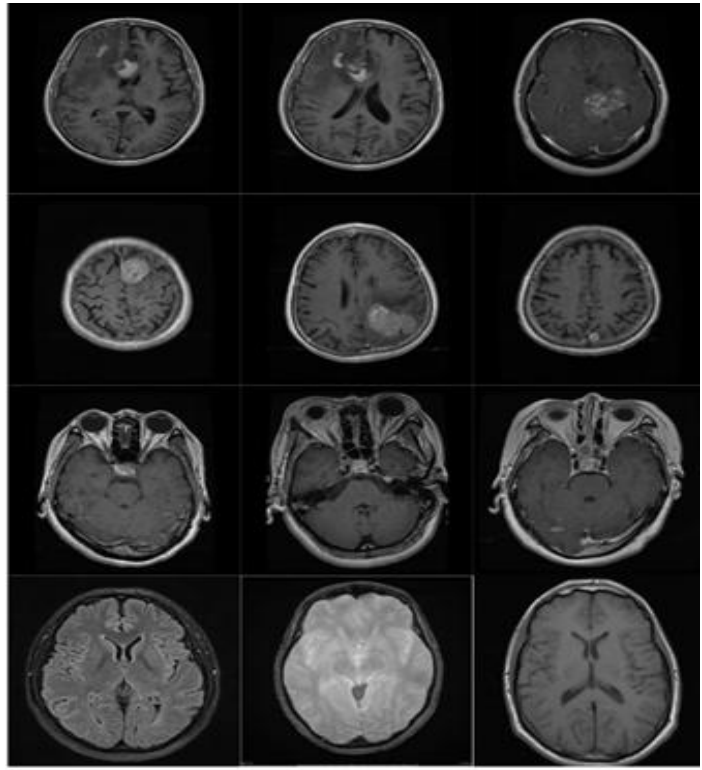

Figure 1. MRI Image Data of Patients with Glioma Brain Tumors (First Row), Meningioma Brain Tumors (Second Row), Pituitary Brain Tumors (Third Row), and Non-Tumor (Fourth Row).

The dataset was partitioned into training and testing subsets, with 80% allocated for training and 20% for testing. Previous studies have demonstrated that this 80:20 ratio optimizes accuracy in image classification tasks [31]. The total dataset includes 3,264 images (noting a correction from the original 3,160 to align with earlier manuscript details), of which 2,611 were used for training and 653 for testing. For binary classification, images of glioma, meningioma, and pituitary tumors were labeled as "positive" (tumor present), while non-tumor images were labeled as "negative" (tumor absent). This binary dataset was similarly split using the 80:20 ratio.

## 2.2. Modeling

This study employed Quantum Convolutional Neural Networks (QCNNs) for modeling, developing two distinct architectures: a multiclass classification model (four classes: glioma, meningioma, pituitary, non-tumor) and a binary classification model (two classes: tumor vs. non-tumor). Both models were constructed using the TensorFlow Sequential framework and consist of multiple layers: an initial quantum convolutional layer, followed by a flattening layer, and subsequent dense layers for classification.

Quantum gates are basic operations that manipulate the state of qubits in quantum computing analogous to logic gates in classical computers that are unitary and reversible. Controlled-NOT (CX or CNOT) and Controlled-Z (CZ) are quantum gates that are frequently used. These CX and CZ are represented in matrix form as in Equations (1) and (2) below.

$$CX = \begin{pmatrix} 1 & 0 & 0 & 0 \\ 0 & 1 & 0 & 0 \\ 0 & 0 & 0 & 1 \\ 0 & 0 & 1 & 0 \end{pmatrix} \qquad (1)$$

$$CZ = \begin{pmatrix} 1 & 0 & 0 & -0 \\ 0 & 1 & 0 & -0 \\ 0 & 0 & 1 & -0 \\ 0 & 0 & 0 & -1 \end{pmatrix} \qquad (2)$$

This combination of CX or CZ can be used to build complex quantum circuits to run the quantum algorithms in this study. The quantum convolutional layer, configured with a filter size of 2, depth of 4, and 4 qubits, leverages controlled-NOT (CX) and controlled-Z (CZ) gates to perform convolution operations. The four filters used are convolution filters consisting of 3 classical convolution filters and 1 quantum convolution filter (the last filter) with varying filter sizes. For the binary classification model, this layer employs a ReLU activation function, followed by a final dense layer with a sigmoid activation function to output binary predictions (0 or 1). Training utilized the binary cross-entropy loss function over 10 epochs. In contrast, the multiclass classification model features a quantum convolutional layer with a ReLU activation function, culminating in a dense layer with a softmax activation function to generate probability distributions across the four classes. This model was trained using the sparse categorical cross-entropy loss function, also over 10 epochs.

## 2.3. Evaluation

Model performance was evaluated using metrics derived from the confusion matrix, including accuracy, sensitivity, specificity, precision, and F1-score. The confusion matrix (Figure 2) quantifies classification outcomes through true positives (TP), false positives (FP), true negatives (TN), and false negatives (FN), providing a robust framework for assessing model efficacy.

Figure 2. Confusion matrix illustrating TP, FP, TN, and FN values

The confusion matrix can be used for accuracy, sensitivity, precision, and F1-score (Figure 2). Accuracy is the rate of correctly predicted positive cases over the total number of cases. Sensitivity refers to the rate of correctly predicted positive cases over total positive cases. Precision means the rate of correctly predicted positive cases from all predicted positive cases. The F1-score balances sensitivity and precision and represents a balanced measure. These metrics establish evaluative benchmarks for the model's effectiveness at generating predictions.

$$Accuracy = \frac{TP + TN}{TP + TN + FP + FN}$$

$$Sensitivity = \frac{TP}{TP + FN}$$

$$Precision = \frac{TP}{TP + FP}$$

$$F_1\ score = 2\frac{Sensitivity \times Precision}{Sensitivity + Precision}$$

## 3. Results and discussion

**Preprocessing Data**

The dataset comprises 3,264 brain MRI scans, including cases diagnosed with glioma, meningioma, and pituitary tumors, as well as non-tumor controls, all resized to 12×12 pixels. This dataset was split into training and testing subsets in an 80:20 ratio, yielding 2,611 training images and 653 testing images. Table 1 details the distribution across the four multiclass categories (glioma, meningioma, pituitary, non-tumor), while Table 2 presents the binary classification split, combining tumor classes (glioma, meningioma, pituitary) as "positive" (tumor present) and non-tumor cases as "negative" (tumor absent).

Table 1 Data splitting to 4 classes (multiclass)

| No | Diagnosis | Train (images) | Validation (images) | Total (images) |
|---|---|---|---|---|
| 1 | Glioma | 712 | 214 | 926 |
| 2 | Meningioma | 755 | 182 | 937 |
| 3 | Pituitary | 728 | 173 | 901 |
| 4 | Brain tumor negative | 416 | 84 | 500 |
|  | Total | 2611 | 653 | 3264 |

Table 2 Data splitting to 2 classes (binary)

| No | Diagnosis | Train (images) | Validation (images) | Total (images) |
|---|---|---|---|---|
| 1 | Brain tumor positive | 2214 | 550 | 2764 |
| 2 | Brain tumor negative | 397 | 103 | 500 |
|  | Total | 2611 | 653 | 3264 |

Table 2 shows that the dataset is not equally distributed regarding classes. This unequal data split may cause the machine learning models to assign different importance to the classes. As a result, the performance would be abysmal for the classes having fewer data samples. Oversampling techniques are utilized to handle this problem by balancing the dataset by augmenting the minority class samples through synthetic data or duplicating the existing samples [32]. This study used Random Oversampling, in which data from the minority class was copied until it was equal in size to the biggest class. The oversampling technique was only performed on the binary dataset because the number of positive and negative data was very high (positive: 2214 and negative: 397). Table 3 shows the binary dataset after oversampling. Oversampling was an integral part of this study to balance the dataset, especially the non-tumour class that initially had fewer samples than the tumour classes.

Table 3 Data splitting to 2 classes after oversampling

| No | Diagnosis | Training data before oversampling | Training data after oversampling | Validation data (images) |
|----|-----------|-----------------------------------|----------------------------------|--------------------------|
| 1 | Brain tumor positive | 2214 | 2214 | 550 |
| 2 | Brain tumor negative | 397 | 2214 | 103 |
| Total | | 2611 | 4428 | 653 |

**Modeling**

The quantum convolutional neural network (QCNN) architecture includes a quantum convolution layer to perform convolution operations. The quantum convolution layer's configuration depends on specific parameters, including filter size and depth. This study used a filter size of 2 and a depth of 4. Filter size dictates the number of qubits needed, which is given by $qubit = filter\_size^2$ in order to affect the input encoding. Input images are encoded into qubits based on the filter size and repeated iteratively until it covers all image pixels. Convolution operations use quantum circuits built from controlled-Not (CX) and controlled-Z (CZ) quantum logic gates. These operations are performed sequentially according to the defined depth, and this paper conducts four convolutions on each image. The quantum convolution layer replaces the classical convolution layer in the classification model.

The QCNN was implemented using TensorFlow's Sequential API, comprising four layers: (1) a quantum convolutional layer with a ReLU activation function, (2) a flattening layer for dimensionality reduction, (3) a dense layer with a ReLU activation function to enhance feature processing, and (4) a final dense layer with either a softmax activation function (multiclass) or sigmoid activation function (binary). Two models were developed: a multiclass model classifying four categories and a binary model distinguishing tumor presence.

**Multiclass model**

In this step—there exist four such categories, supposed to point out either a glioma tumour, meningioma tumour, pituitary tumour, or no presence of a tumour—coming out from the second dense layer is the probability distribution over those classes, given by applying the Softmax function; the class having the highest probability is then adopted as an outcome. Figure 3 provided a visual representation of a sequential neural network model architecture for multiclass classification, specifically designed for a task involving convolutional layers followed by dense (fully connected) layers.

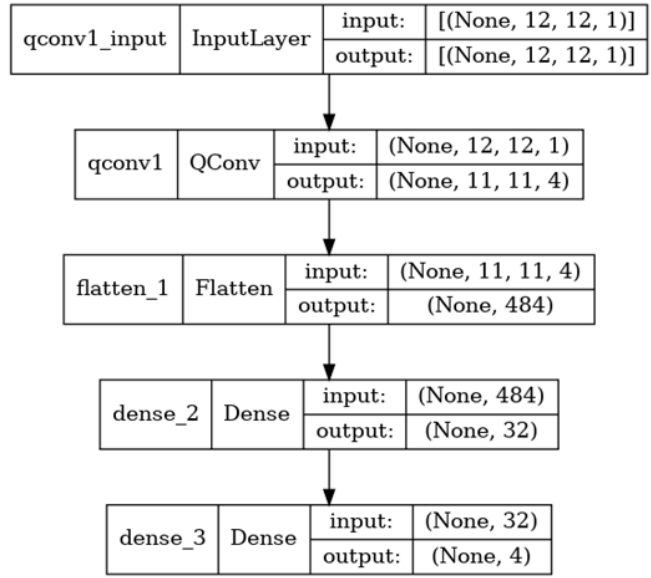

Figure 3. Visualization of brain tumor image multiclass classification model

**Binary model**

In binary classification, the model predicts whether an input image belongs to either the tumour-positive or tumour-negative classes. The final output from the second dense layer brings out a value of either 0 or 1 for the two classes under consideration using a Sigmoid activation function. Figure 3 illustrates the binary classification process. This process is the same as in the multiclass model, except for the output produced by the last dense layer. In this case, the output can be either 1 or 0, representing whether the input image is classified as positive for a brain tumour or negative for a brain tumour.

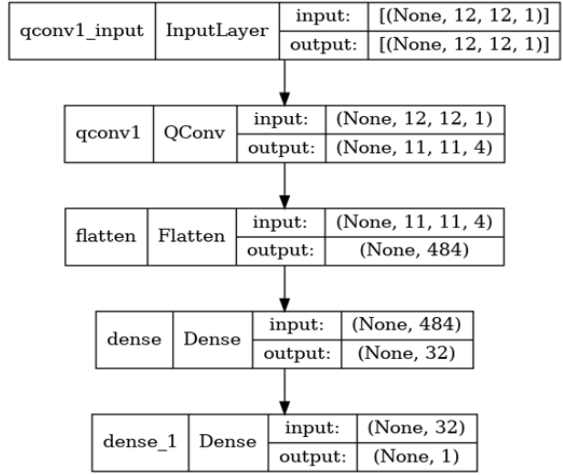

Figure 4. Visualization of brain tumor image binary classification model

**Multiclass classification**

The multiclass model was trained on the four-class dataset, producing probability distributions for each class. Figure 5 plots training epochs against accuracy, showing a consistent increase in accuracy and a corresponding decrease in loss, indicating effective learning. The original imbalanced dataset yielded an accuracy of 52%, precision of 53%, sensitivity of 49%, and F1-score of 50%. Post-oversampling (applied in a separate experiment to assess impact, though not standard for multiclass here), accuracy improved to 62%, with precision at 65%, sensitivity at 62%, and F1-score at 63%, reflecting a 10% validation accuracy gain and 20% training accuracy improvement.

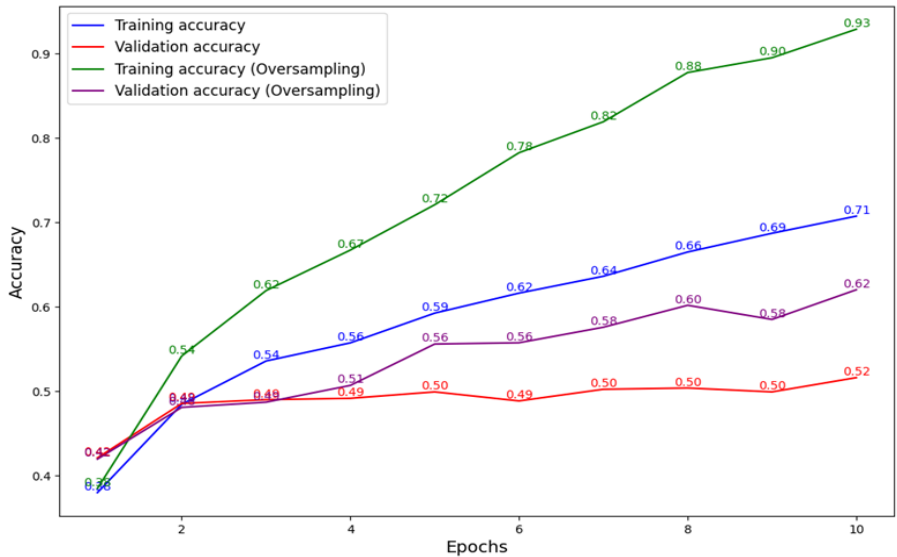

Figure 5. Graph of epoch against multiclass model accuracy

Figure 5 show the epoch-versus-accuracy graph, accuracy improve with the rise of epoch. Comparing both graphs portrays opposite trends, proving that the model becomes more accurate as training progresses. The loss function decreased every epoch, indicating that the model was good at classification. These results show that the oversampling technique can improve the accuracy of the model by 20% in the data training process and 10% in the validation process. The model was evaluated with standard metrics: accuracy, precision, sensitivity, and F1-score. The model trained on the original imbalanced dataset only achieved an accuracy of 52%, sensitivity of 49%, and F1-score equivalent to 50%. Conversely, the model trained on the oversampled dataset improved to 62% in accuracy, 65% in precision, 62% in sensitivity, and 63% in F1-score.

**Binary classification model**

The binary classification model was developed to classify brain MRI images into tumour-positive and tumour-negative categories. The accompanying figures illustrate the epoch-versus-accuracy and epoch-versus-loss graphs during the training of the binary classification model. The training was conducted on a prepared 2-class dataset, utilizing all processes across the model's layers to produce an output value of 0 or 1.

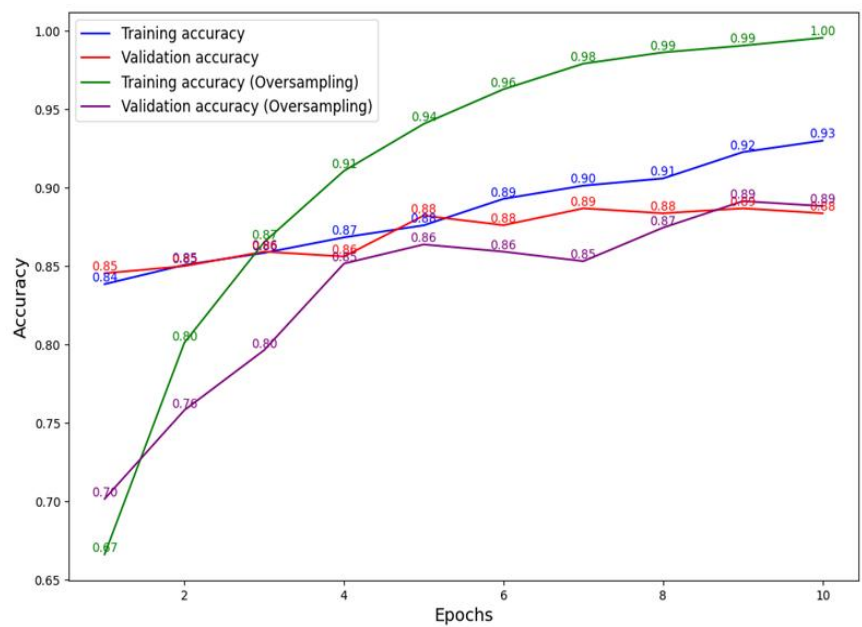

Figure 6 Graph of epoch against binary model accuracy

Figure 6 shows the epoch versus accuracy graph; it reflects an improvement in accuracy over time, demonstrating increased reliability in classification by the model. The binary model also uses the same metrics for the evaluation as multiclass classification. The model trained on the original dataset resulted in 88% accuracy, 83% precision, 67% sensitivity, and an F1-score of 72%. After oversampling, the model was slightly better, with 89% accuracy, 79% precision, 80% sensitivity, and an F1-score of 80%.

Table 4 Evaluation result of validation data

|  | Accuracy | Precision | Sensitivity | F1-score |
| --- | --- | --- | --- | --- |
| Multiclass classification | 52% | 53% | 49% | 50% |
| Multiclass classification with oversampling | 62% | 65% | 62% | 63% |
| Binary classification | 88% | 83% | 67% | 72% |
| Binary classification with oversampling | 89% | 79% | 80% | 80% |

Table 4 summarizes the performance of all models. The QCNN exhibited strong binary classification performance, achieving up to 89% accuracy with oversampling, but struggled with multiclass classification, peaking at 62%. This disparity suggests that while QCNNs excel in simpler binary tasks, their efficacy diminishes with increased class complexity. Potential limitations include the dataset's modest size (3,264 images), the intricate feature variability across four classes, and the quantum circuit's design—relying on CX and CZ gates within a single convolutional layer—which may lack the sophistication needed for multiclass differentiation.

Datasets are the most crucial element in the development and assessment of CNN and QCNN performance in image classification. These architectures cannot learn well and optimally without quality datasets. The image dataset used in this study is very limited and not varied (only sourced from some patient data) which causes a non-optimal model that only focuses on details that are not directly related to the most influential features in distinguishing classes. The imbalance data in each class is also an important factor to avoid bias occurring in the majority dataset, where the model is forced to learn this majority dataset and vice versa in the minority data. One way to overcome dataset limitations is the application of oversampling techniques. This technique significantly enhanced binary performance by addressing class imbalance, though its multiclass impact was less pronounced due to the dataset's inherent balance. These findings highlight QCNNs' potential in medical imaging, particularly for binary tasks, but indicate a need for advanced quantum architectures to tackle multiclass challenges effectively.

To further interpret these findings, it is important to emphasize that this study applies a compact four-qubit QCNN directly to a real-world brain MRI dataset rather than relying on simple benchmark datasets such as MNIST, which remain the focus of most prior QCNN research. Moreover, the direct performance comparison with a lightweight classical CNN under the same data split and preprocessing settings provides a practical perspective on how quantum models currently perform in a clinical imaging context. While the QCNN shows promising results for binary classification with a significantly smaller number of trainable parameters, its limited multiclass performance suggests that the current quantum circuit design—specifically the fixed filter size, shallow depth, and simple entanglement using only CX and CZ gates—may constrain the model's ability to capture subtle differences among similar tumor types. Although this study did not include multiple circuit design experiments due to hardware and simulator limitations, we acknowledge that circuit depth, qubit connectivity, and gate sequence are critical factors influencing the model's expressiveness and resilience to noise. Therefore, future work will focus on systematically exploring alternative quantum circuit architectures, deeper entanglement topologies, and variational re-uploading techniques to enhance feature extraction capacity while mitigating the effects of quantum decoherence. Combining such circuit optimizations with larger and more diverse training datasets is expected to improve multiclass classification performance. This line of research is a crucial step towards demonstrating practical quantum advantage for complex medical imaging tasks on near-term quantum devices. Overall, the results support the feasibility of deploying QCNNs for automated brain MRI screening in binary scenarios, providing comparable performance to classical CNNs but with far fewer parameters and shorter training time in simulation. These insights lay a foundation for further bridging the gap between theoretical quantum models and realistic, clinically meaningful applications in neuro-oncology.

## 4. Conclusion

This study demonstrates that the Quantum Convolutional Neural Network (QCNN) model is applicable to both multiclass and binary classification of brain MRI images. In the multiclass scenario, the model classified images into glioma, meningioma, pituitary, and non-tumor categories, achieving an accuracy of 52% with the original dataset. Balancing the data through oversampling enhanced its performance, increasing accuracy to 62%. For binary classification, the model distinguished tumor-positive from tumor-negative images with an accuracy of 88%, precision of 83%, sensitivity of 67%, and F1-score of 72% on the original dataset. Following oversampling to address class imbalance, performance improved to 89% accuracy, 79% precision, 80% sensitivity, and an F1-score of 80%. While the QCNN exhibits robust accuracy in binary classification, its multiclass performance remains comparatively modest. Several factors may contribute to this disparity: the dataset's limited size (3,264 images) and inherent complexity, the challenge of distinguishing four distinct classes, the potentially suboptimal configuration of quantum circuits using controlled-NOT (CX) and controlled-Z (CZ) gates for multiclass tasks, and the simplicity of the model architecture, which relies on a single quantum convolutional layer. These limitations suggest that while QCNNs hold promise for binary medical imaging applications, further refinements—such as expanded datasets, optimized quantum circuits, or deeper architectures—are necessary to enhance multiclass classification efficacy.